\documentclass[twocolumn,aps,prd]{revtex4}
\usepackage{graphicx}
\usepackage{natbib}

\newcommand{\etal}{{et~al.}}
\def\aa{{\sl Astron.\ \&\ Astrophys. \ }}

\def\apj{{\sl Astrophys.\ J. \ }}
\def\apjl{{\sl Astrophys.\ J.\ Lett. \ }}
\def\apjs{{\sl Astrophys.\ J.\ Supp. \ }}
\def\araa{{\sl Ann.\ Rev.\ Astron.\ Astrophys. \ }}

\def\mnras{{\sl MNRAS \ }}
\def\mnrasl{{\sl MNRAS \ Lett. \ }}

\def\np{{\sl Nucl.\ Phys. \ }}

\def\pr{{\sl Phys.\ Rep. \ }}
\def\prd{{\sl Phys.\ Rev.\ D \ }}
\def\prl{{\sl Phys.\ Rev.\ Lett. \ }}

\def\ptps{{Prog.\ Theor.\ Phys.\ Suppl. \ }}
\def\rmp{{\sl Rev.\ Mod.\ Phys. \ }}

\newcommand{\ie}{{i.e.}}

\newcommand{\lsim}{\,\lower2truept\hbox{${<\atop\hbox{\raise4truept\hbox{$\sim$}}}$}\,}
\newcommand{\gsim}{\,\lower2truept\hbox{${>\atop\hbox{\raise4truept\hbox{$\sim$}}}$}\,}

\def\ie{{\rm i.e.$\,$}}

\def\etal{{\rm et al.$\,$}}

\newcommand{\be}{\begin{equation}}
\newcommand{\ee}{\end{equation}}
\newcommand{\bea}{\begin{eqnarray}}
\newcommand{\eea}{\end{eqnarray}}

\begin{document}

\title{Dark energy records in lensed cosmic microwave background}

\author{Viviana Acquaviva$^{1,2}$, Carlo Baccigalupi$^{1,2,3}$}
\affiliation{$^{1}$ SISSA/ISAS, Astrophysics Sector, Via Beirut, 4,
I-34014 Trieste, Italy\\
$^{2}$ INFN, Sezione di Trieste, Via Valerio 2,
I-34014 Trieste, Italy\\
$^{3}$ Institut f$\ddot{\it u}$r Theoretische Astrophysik,
Universit$\ddot{\it a}$t Heidelberg, Albert-$\ddot{U}$berle-Strasse 2, D-69120
Heidelberg, Germany}

\begin{abstract}
We consider the weak lensing effect induced by linear 
cosmological perturbations on the cosmic microwave 
background (CMB) polarization anisotropies. 
We find that the amplitude of the lensing peak in 
the BB mode power spectrum is a faithful tracer of the 
dark energy dynamics at the onset of cosmic acceleration. 
This is due to two reasons. First, the lensing power is non-zero 
only at intermediate redshifts between the observer and the source, 
keeping record of the linear perturbation growth rate at the 
corresponding epoch. Second, the BB lensing signal is expected to 
dominate over the other sources. The lensing distortion on the TT 
and EE spectra do exhibit a similar dependence on the dark energy 
dynamics, although those are dominated by primary anisotropies. \\
We investigate and quantify the effect by means of exact tracking 
quintessence models, as well as parameterizing the dark energy equation 
of state in terms of the present value ($w_{0}$) and its asymptotic value in the past ($w_{\infty}$); in the interval allowed by the present 
constraints on dark energy, the variation of $w_{\infty}$ induces a significant change 
in the BB mode lensing amplitude. 
A Fisher matrix analysis, under conservative assumptions concerning 
the increase of the sample variance due to the lensing non-Gaussian statistics, 
shows that a precision of order $10\%$ on both $w_{0}$ and $w_{\infty}$
is achievable by the future experiments probing a large sky area 
with angular resolution and sensitivity appropriate to detect the lensing 
effect on the CMB angular power spectrum; the forecast precision reaches 
a few percent for highly dynamic models whose dark energy abundance 
at the epoch when lensing is most effective is sensibly larger than 
the present one, \ie for $w_\infty \gsim$ -0.5.
These results show that the CMB can probe the differential redshift 
behavior of the dark energy equation of state, beyond its average. 
\end{abstract}

\maketitle   

\section{Introduction}
\label{i}

One of the most challenging issues in modern cosmology is the 
comprehension of the nature of the dark energy, the unknown 
component representing about $70\%$ of the cosmological critical 
density today, responsible for a late phase of acceleration in the 
cosmic expansion (see \cite{padmanabhan_2003,peebles_ratra_2003} 
and references therein). \\
The first report on the evidence of cosmic acceleration was due to
the magnitude-redshift relation 
inferred by type Ia supernovae \cite{perlmutter_etal_1999,riess_etal_1998}; 
the cosmic microwave background (CMB) experiments, combined with the 
data on large scale structure, confirmed and strengthened that 
result (see \cite{spergel_etal_2003} and references therein). \\
The simplest interpretation for the acceleration can be given in terms of a 
Cosmological Constant, leading however to serious theoretical problems 
concerning its magnitude. In fact, its energy scale is required to be 
123 orders of magnitude lower than the Planck energy, possibly the only 
one relevant in the very early universe; and its value has to be such to 
render it comparable with the matter density at the present epoch. These 
issues are known as fine-tuning and coincidence, respectively (see e.g. 
\cite{peebles_ratra_2003}). \\ 
The concept of dark energy generalizes the Cosmological Constant, 
allowing dynamics of the equation of state and fluctuations of the 
dark energy component, in the attempt to alleviate these
tweaking problems and to find clues to unveil the physical mechanism 
giving rise to the acceleration. \\
This is the case of the Quintessence, a self-interacting scalar 
field evolving according to different potential energies. For reference, 
the potentials which have been studied extensively in the literature 
are characterized by an exponential shape \cite{wetterich_1988}, inverse 
power law \cite{ratra_peebles_1988}, or a combination of those 
\cite{brax_martin_2000}, and have been suggested in the context of 
particle physics beyond the standard model. \\
The present measurements are consistent with the Cosmological Constant case, 
$w=-1$ \cite{seljak_etal_2006}, with a precision of about ten percent. 
However, these observations do not have the capability to address the 
dynamics of the dark energy yet: the corresponding constraints either 
concern models where the dark energy is constant, or may be interpreted 
as constraints on the redshift average of the equation of state. \\
Most dark energy models, including the ones quoted above, may account for 
a present equation of state close to the Cosmological Constant case, but 
have a significantly different evolution at the epoch of equality between 
dark matter and energy, occurring at redshift $z\simeq 0.5$. 
Therefore, on the basis of the Cosmological Constant problems, and having 
no other theoretical clue about the nature of the dark energy, it is clearly 
crucial to investigate its behavior at the onset of acceleration,   
where the models most differ. \\
The dark energy dynamics is and will be probed in the first place by the 
same observables which gave evidence for cosmic acceleration. 
The latest measurements of type Ia supernovae measurements are from 
space, from the Hubble Space Telescope (HST), reaching a redshift of 
about 0.5 \cite{chiocchiatti_etal_2006}. With appropriate experimental 
resources, the type Ia supernovae measurements may get farther in redshift, 
collecting data from events occurred before the onset of acceleration, 
up to a maximum redshift between 1.5 and 2, thus probing the whole redshift 
interval where the dark energy is relevant \cite{linder_etal_2005}. 
At the same time, the combination of CMB and large scale structure 
probes the dark energy, through the modified history of structure 
formation. The local large scale structure is probed directly by 
the spatial distribution of galaxies. Indications on the cosmological 
structures at higher redshifts come from the distribution of the 
Ly$\alpha$ clouds \cite{viel_etal_2006}. The imprints of the baryon acoustic 
oscillations in the dark matter distribution have been observed 
\cite{eisenstein_etal_2005}. In addition, the CMB undergoes a geometrical 
shift of the acoustic peak locations in the angular domain, because of the 
modified distance to the last scattering resulting from the change 
in the expansion rate. The 2dF survey \cite{cole_etal_2005} and Sloan Digital 
Sky Survey (SDSS, see \cite{seljak_etal_2005} and references therein), and the 
CMB experiments culminating with the Wilkinson Microwave Anisotropy 
(WMAP, see \cite{spergel_etal_2003} and references therein) contributed 
to set the quoted constraint on the dark energy equation of state 
\cite{seljak_etal_2006}. 
On the other hand, as we stressed above, almost all those observables with 
the only exception perhaps of the baryon acoustic oscillations, rely on 
a line of sight integral of the light rays, picking up comparable 
contributions at all epochs from emission through observation; for those 
observables, the redshift behavior of the dark energy is averaged, and 
in particular markedly influenced by the expansion at recent epochs, where the 
observations require the dark energy to be close to a Cosmological Constant. 
In particular, the large scale structure survey data are mostly determined by nearby 
structures, only extending up to a redshift of about 0.1, while the CMB projection 
effect is dominated by the effect of the dark energy when it is most relevant, 
i.e. at the present. Therefore it is also important to study observables 
capable to pick up the dark energy abundance at the onset of acceleration. 
The lensing effect is a unique tool for this purpose, 
and is the subject of the present work. The reason is an elementary geometric 
property of lensing, yielding a null cross section if the lens position 
coincides with the observer or the source, thus probing intermediate regions 
only (see \cite{bartelmann_schneider_2001} for reviews). \\
The weak lensing in cosmology, i.e. the large scale shear injected 
on the background light by forming cosmological structures, is 
one of the most promising observables for future studies on dark 
matter and energy; a great effort is directed towards the study of 
the ellipticity induced by weak lensing on distant galaxies in 
the optical band, and how that is sensitive to the dark energy 
properties and other cosmological parameters. 
In principle, by measuring the weak lensing shear induced on shells of 
galaxies at different redshifts, either statistically or looking 
at the lensing induced by one single large structure, one has information 
on the redshift evolution of the dark energy abundance, which is equivalent 
to the knowledge of the equation of state dynamics 
\cite{hu_jain_2003,jain_taylor_2003,hu_2002}. 
In this perspective, a large experimental effort is ongoing; 
see \cite{refregier_2003} 
for reviews on the existing observations and future projects, 
\cite{SNAP,Yeche_et_al,song_knox_2004} 
for parameter forecasts, also in connection with the constraints 
expected from other cosmological probes. \\
In particular, if the source can be considered at infinity the 
lensing cross section is non-zero at redshifts roughly 
between 0.1 and 10, peaking at $z \simeq 1$, rather independently of the 
particular cosmological model considered, and thus most relevant 
to study the universe at the corresponding epoch. 
This is the case of this work, where we consider the lensing of the CMB 
\cite{lewis_chal,hu_2001,hu_2000,goldberg_spergel_1999,zaldarriaga_seljak_1998}. 
The relevance of CMB lensing as a tool for constraining the dark 
energy has been investigated for what concerns the statistics 
of order higher than the second for the anisotropies in total intensity 
\cite{verde_spergel_2002,giovi_etal_2005}. Indeed, even if the primordial CMB 
anisotropies obey a perfect Gaussian distribution, after lensing their 
statistics is modified as a result of the correlation among different scales 
induced by the lensing itself, in total intensity and polarization. 
Such effect appears already at the level of anisotropies in the total intensity, 
which represent the strongest component, and that is the reason why it has been 
studied so far mainly in that respect. As the present subject concerns the CMB 
lensing and is therefore related to that, we return on this issue in the 
concluding remarks. \\
Here we focus on a different domain where the lensing is relevant for 
CMB, and precisely the anisotropy angular power spectrum, in particular 
for the the curl or BB component in the CMB polarization signal. 
The lensing re-distributes the primordial power and correlates different 
scales; as a result, the acoustic peaks in the total intensity (TT), 
gradient (EE) components of the CMB polarization and their correlation (TE) 
are smoothed, and some power is transferred from such scales to the damping 
tail. Moreover, a central aspect of the present 
work is that the gradient component of CMB polarization leaks 
into the BB modes, causing a broad peak centered on the angular scales 
of a few arcminutes, roughly corresponding to a multipole $ l \simeq 1000$. 
Although the lowest expected contribution to the CMB anisotropies, 
this observable is entirely caused by lensing, and basically unbiased 
by primordial power; a contamination due to primordial gravitational waves 
may arise on larger angular scales, multipoles of about 100, where however 
the lensing signal is rapidly decreasing approaching the 
super-horizon regime. Furthermore, as we stated already, 
the lensing is a non-Gaussian process, correlating cosmological 
perturbations on different angular scales, while the primordial tensor power 
is expected to be close to Gaussianity. This difference might be crucial to 
deconvolve the two patterns \cite{hirata_seljak_2003}. The non-Gaussian 
distribution of lensing has an impact already at the level of the 
variance in the angular power spectrum of the BB modes, which is currently 
under investigation \cite{lewis_2005,smith_etal_2004,kaplinghat_etal_2003}.\\
Therefore, since the lensing cross section is non-zero at intermediate 
redshifts only, as specified above, we do expect some relevance in studying 
this effect for investigating the dark energy at non-zero redshifts; in particular, 
being the BB power on the arcminute scale caused by lensing only, this 
relevance should be directly reflected by the behavior of that component. 
An analogous study, pointing out the lensing cross section redshift distribution 
and focusing on the high redshift dark energy behavior, has been performed 
considering the CMB third order statistics in total intensity anisotropies 
\cite{giovi_etal_2005}. 

Our treatment is based on a previous work \cite{acquaviva_etal_2004}, 
casting the cosmological weak lensing theory in the context of 
scalar field dark energy models with arbitrary kinetic and potential 
forms in the fundamental Lagrangian, and also coupling arbitrarily to the 
Ricci scalar. 
The aim of that work was to embrace the most general scalar field 
dark energy models, including as a particular case the minimally coupled, 
purely self-interacting Quintessence, which is the scenario upon which 
this work is based. In section \ref{wlade} we recall the relevant issues 
concerning the lensing computation in dark energy cosmologies. In section 
\ref{lcmbpps} we derive and discuss the lensed CMB power spectra and their 
dependence on the dynamical dark energy properties. In section \ref{fma} 
we evaluate the impact on a parametric analysis of CMB data. Finally, in 
section \ref{c} we draw our conclusions. 

\section{Weak lensing and dark energy}
\label{wlade}

In this section we describe the cosmological models we consider 
throughout this paper, and we outline the physics of lensing on the CMB 
total intensity and polarization anisotropies; for more details, see 
\cite{acquaviva_etal_2004} and references therein. We focus on the 
modifications to the Boltzmann codes which numerically evolve 
cosmological perturbations required in order to take into account the 
effects of lensing in scalar field dark energy cosmologies. 

\subsection{Dark energy cosmology}
\label{dec}
In this paper we will consider the tracking Quintessence scenarios, where the 
dark energy is described through a scalar field $\phi$ (see e.g. 
\cite{peebles_ratra_2003}). The associated action is of the form:
\bea
\label{quintaction}
S&=&\int d^4x \sqrt{-g}\cdot\nonumber\\
&\cdot&\left[ \frac{1}{2\kappa}R-
\phi^{;\mu}\phi_{;\mu}\frac{1}{2} -
V(\phi) + {\cal{L}}_{\rm{fluid}}\right]\ .
\eea
We chose two representative 
models where the equation of state has a mild and violent redshift behavior, 
respectively for the inverse power law potentials, (IPL \cite{ratra_peebles_1988}) 
and those inspired by super-gravity theories (SUGRA \cite{brax_martin_2000}): 
\begin{equation}
V(\phi )=\frac{M^{4+\alpha}}{\phi^{\alpha}}\ \ ,\ \ 
V(\phi )=\left(\frac{M^{4+\alpha}}{\phi^{\alpha}}\right)
e^{4\pi G\phi^{2}}\ .
\end{equation}
These scalar field dark energy models are implemented and integrated exactly, also 
considering Quintessence fluctuations, using DEfast, a modification of the Boltzmann code 
for numerical integration of cosmological background and linear perturbations 
based on the version 4.0 of CMBfast \cite{seljak_zaldarriaga_1996}, which has been 
used in several papers, see \cite{baccigalupi_etal_2000,perrotta_baccigalupi_1999} 
and references therein. \\
For what concerns the background evolution, a variety of models including the 
ones above are well described by essentially two parameters: those are 
the present value $w_{0}$ of the equation of state and its first
derivative with respect to the scale factor $a$, $-w_{a}$ \cite{cheva_pol,linder_2003}. 
In this framework, the evolution of the equation of state with the scale factor can be 
written as
\begin{equation}
\label{linder_parametrization}
w(a)=w_{0}+w_{a}(1-a)=w_{\infty}+(w_{0}-w_{\infty})a\ ,
\end{equation}
where $w_{\infty}$ is the asymptotic value of $w$ in the past. We will exploit 
the parameterization above in Section \ref{fma}, in order to evaluate the precision 
achievable on the measure of $w_0$ and $w_{\infty}$ from the CMB total intensity 
and polarization angular power spectra. 

\subsection{Weak lensing and Boltzmann numerical codes in cosmology}
\label{wlabncic}

The effect of gravitational lensing on the CMB spectra had been first introduced 
in the CMBfast code by Zaldarriaga and Seljak \cite{zaldarriaga_seljak_1998} for 
Cold Dark Matter (CDM) cosmologies including a Cosmological Constant ($\Lambda$CDM). 
In their formalism the effect on the correlation functions can be understood as 
the convolution of the unlensed spectra with a Gaussian filter determined by the 
{\it lensing potential} \cite{bartelmann_schneider_2001}.
The expression for the lensing potential has to be generalized in scalar-tensor 
cosmologies because of the presence of anisotropic stress already 
at a linear level \cite{acquaviva_etal_2004}. In the present scenario, however, 
this is not required and the structure of the quantities relevant for computing 
the lensing effect is formally unchanged. \\
Therefore, following the notation of the original paper \cite{zaldarriaga_seljak_1998}, 
and showing for simplicity the TT case only, the anisotropy correlation between 
directions ${\bf \theta}$ and ${\bf \theta}^\prime$, deflected by $\delta{\bf \theta}$ 
and $\delta{\bf \theta}^\prime$, respectively, and separated by an angle $\theta$ 
is expanded in the harmonic space exploiting the flat sky approximation, taking 
the form 
\be
C_{TT}(\theta) = \int \frac{d^2 {\bf l}}{(2\pi)^2} e^{i l \theta \cos{\phi_l}} 
\langle e^{i{\bf l}\cdot(\delta {\bf \theta} - \delta {\bf \theta}^\prime)}
\rangle C_{\tilde{TT}l} 
\ee
where the expectation value in the above equation represents the ensamble average 
and is expressed as
\be
\langle e^{i{\bf l}\cdot(\delta {\bf \theta} - \delta {\bf \theta}^\prime)}\rangle = 
e^{- \frac{l^2}{2}[\sigma_0^2(\theta) + \cos{(2 \phi_{\bf l})} \sigma^2_2(\theta)]},
\ee
and the quantities $\sigma_0^2(\theta)$ and $ \sigma_2^2(\theta)$ are given by
\bea
\label{sigma0}
\sigma^2_0(\theta) &=& 16\pi^2
\int^{\chi_{\text{rec}}}_0 
W^2(\chi,\chi_{\text{rec}})d\chi
\int^\infty_0 
k^3 dk \cdot \nonumber\\
&\cdot& P_\Phi(k,\tau = \tau_0 - \chi) 
[1 - J_0(k\theta\chi)]\ ,
\eea
and
\bea
\label{sigma2}
\sigma^2_2(\theta) &=& 16\pi^2
\int^{\chi_{\text{rec}}}_0 
W^2(\chi,\chi_{\text{rec}})d\chi
\int^\infty_0 k^3 dk \cdot \nonumber\\
&\cdot& P_\Phi(k,\tau = \tau_0 - \chi)
J_2(k\theta\chi)\ .
\eea
Here $k$ is the wavenumber, $J_{l}$ is the Bessel function of order 
$l$, $\chi$ is the comoving radial distance, $\tau$ is the cosmological conformal time,  
$P_\Phi$ is the power spectrum of the gravitational potential, and $W$ is a function 
accounting for the cosmic curvature, which amounts to $1-\chi/\chi_{\text{rec}}$ 
for a flat universe. In $\Lambda$CDM cosmologies and most of the numerical codes dealing 
with them, including CMBfast, the quantities appearing in the above equations may be 
computed independently of the main routine which performs the integration of the 
hierarchical Boltzmann equations. 
This can be done because the power spectrum of matter density perturbations $\Delta_{m}$ 
can be factorized in two terms, depending respectively only on the wavenumber and the 
redshift: 
\be
P_{\Delta_{m}}(k,\chi) = Ak^{n}\cdot T^2(k,0)g^2(\chi)\ .
\ee
$Ak^{n}$ here represents the primordial power, $T$ is the transfer function of 
density perturbations taking into account the evolution on sub-horizon scales, 
and $g$ is the perturbation linear growth factor. 
However, this separation is only convenient if one is provided with a satisfactory 
analytical fit of the growth of perturbations, which is not the case unless we ignore the 
influence of Quintessence perturbations, which do make a non-negligible effect on 
large scales \cite{ma_etal_1999}. 
To account for these changes, we evaluate numerically the gauge invariant expression 
of density perturbations from all fluctuating components 
$\Delta$ \cite{kodama_sasaki_1984}. This quantity
is computed and saved while the main routine performs the integration of the hierarchical 
Boltzmann equations, and used later for the numerical integration of the quantities 
(\ref{sigma0}) and (\ref{sigma2}), which include all fluctuating components. \\ 
A separate issue concerns the normalization constant $A$ above when lensing 
is taken into account. The lensed perturbation spectra no longer depend 
linearly on the primordial normalization, since the lensing is a second 
order effect, being sourced by cosmological structures and acting on CMB anisotropies. 
Consequently, the description of the primordial anisotropy power has 
to be treated appropriately to take into account this occurrence; we modified 
our Boltzmann code in order to require the primordial normalization as an input, 
which is among the cosmological parameters to be constrained in Section \ref{fma}. \\
This is different from the procedure followed in Section \ref{lcmbpps}, where we investigate 
phenomenologically the impact of lensing 
on the CMB angular power spectra as a function of the dark energy parametrization 
given above. The remaining parameters will be tuned to a fiducial value, 
in order to highlight differences and behavior induced exclusively by the variation of the 
underlying dark energy model; in particular, for the purposes of this Section only, 
the scenarios under examination are set to have the same amplitude of the 
primordial perturbations. An equivalent parametrization of the strength 
of primordial perturbations might be given 
with reference to the present epoch, usually by means of the variance evaluated 
on a scale of $8h^{-1}$ Mpc, $\sigma_8$; we choose the parametrization in terms 
of the primordial amplitude for numerical convenience, and we verify that the 
relevance of the effects we find does not depend on this choice.

\section{Lensed CMB polarization power spectra}
\label{lcmbpps}

\begin{figure}[t]
\centering
\includegraphics[width=\linewidth]{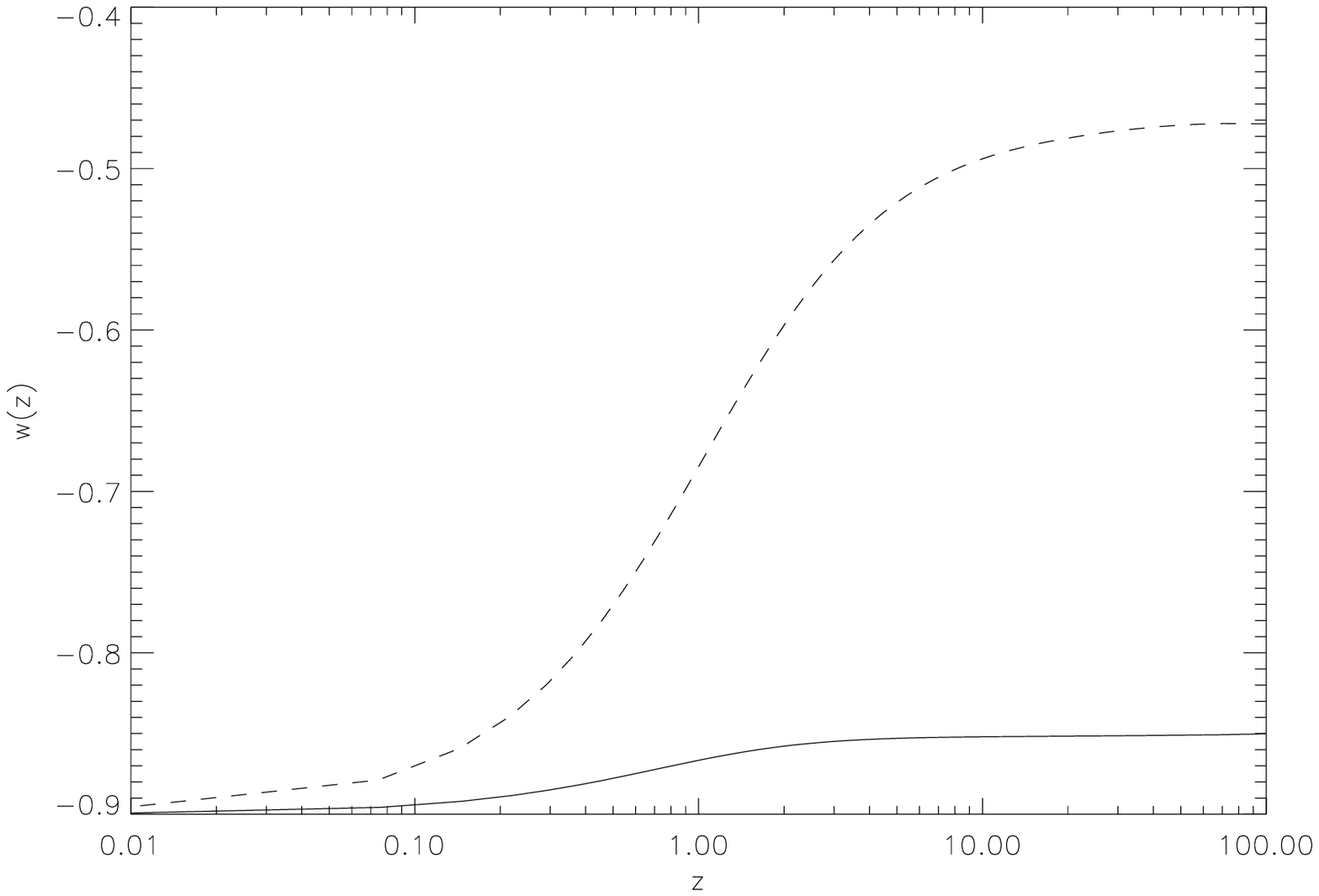} 
\includegraphics[width=\linewidth]{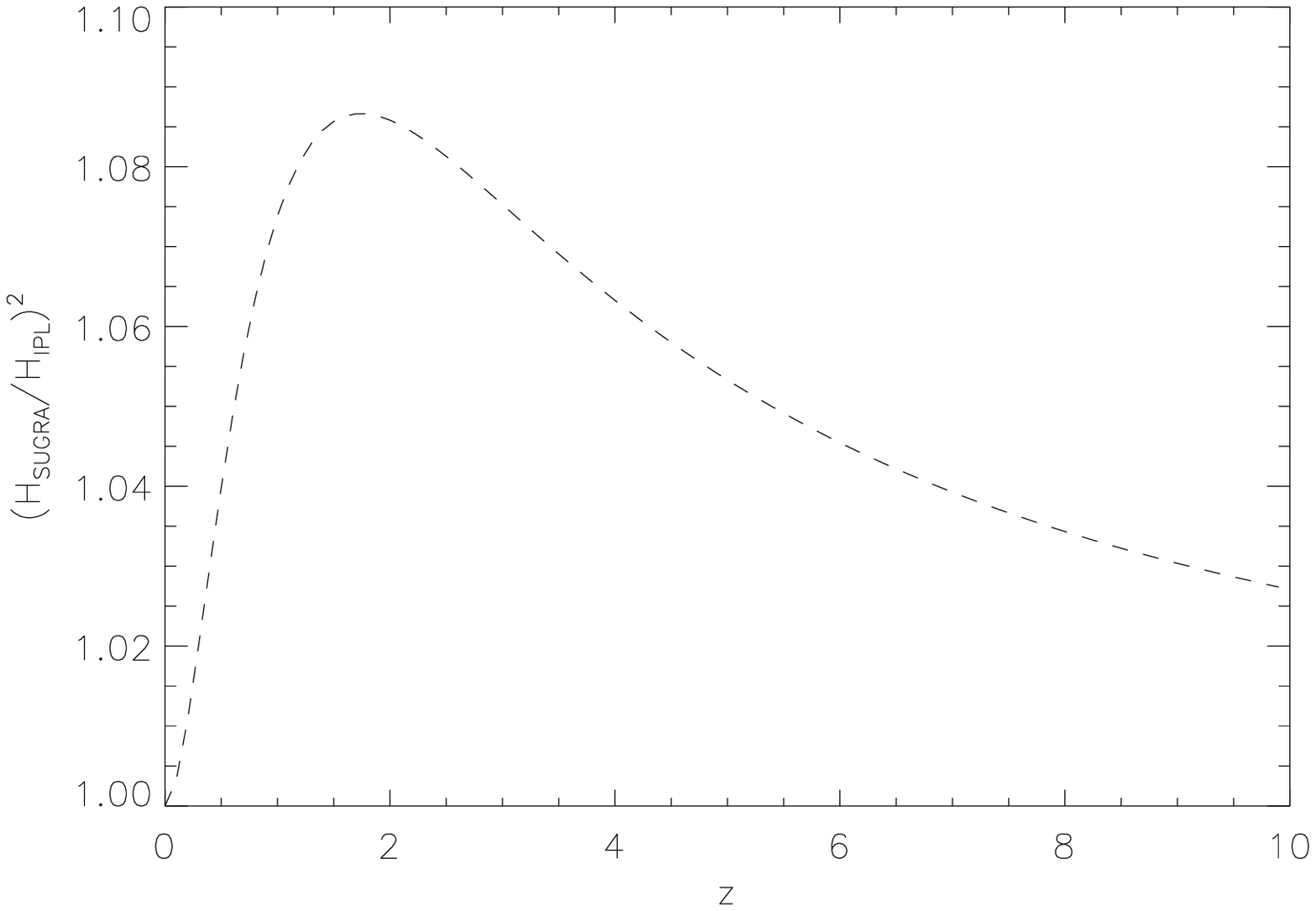}
\caption{Top: evolution of the equation of state of dark energy for the SUGRA (dashed line) 
and IPL (solid line) models. Bottom: ratio of $H^{2}$ in the two models as a function of 
redshift.}
\label{wred}
\end{figure}

\begin{figure}
\includegraphics[width=\linewidth]{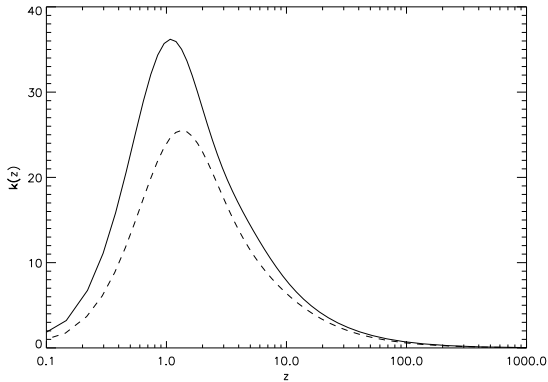} 
\caption{Lensing kernel for $\theta = 4 \cdot 10^{-3} $ rad, 
for the SUGRA (dashed line) and IPL (solid line) models, 
having the same equation of state today.}
\label{fig:geomSI}
\end{figure}

\begin{figure}[t]
\centering
\includegraphics[width=\linewidth]{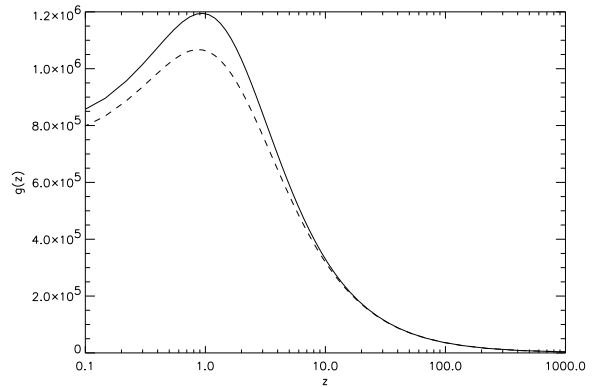} 
\caption{Growth factor of the perturbations for a comoving wavenumber 
$k = 0.1$ Mpc$^{-1}$, for the SUGRA (dashed line) and IPL (solid line) 
models. In the two models, $g$ has the same value at infinity.}
\label{fig:tfdeltaSI}
\end{figure}

We consider the two dark energy models discussed in the previous Section, 
as they well represent the different dynamics that the dark energy might 
have. We study the behavior of the relevant lensing quantities, 
showing results for the corresponding lensed CMB power spectra. In 
particular, we focus on the effect induced by the dark energy behavior 
at the epoch when the lensing power injection is effective. We give 
a qualitative description of how the lensing peak breaks the degeneracy 
between $w_{0}$ and $w_{\infty}$ affecting the TT, TE and EE spectra, 
described later in detail. \\
Both the SUGRA and the IPL models are characterized by two parameters, 
the index of the power law $\alpha$, and the mass of the field $M$. As 
it is well known \cite{liddle_scherrer_1999,steinhardt_etal_1999} 
they both admit attractor trajectories for the field dynamics in the 
early universe, known as tracking solutions. These are characterized by 
a relative independence of the field dynamics on its mass. 
The relevant parameter ruling the motion of the field is $\alpha$, which 
in our example is set respectively as $-2.21$  and $-0.34$ for the SUGRA and 
IPL models. On the other hand, the mass sets the normalization of the dark energy 
density along the trajectory, being therefore crucial to achieve acceleration 
today, and must be set accordingly. \\
First of all we want to discuss qualitatively the effects on the 
corresponding background evolution, where we expect to see the most relevant 
differences between the two models; the key point of the comparison is the 
behavior of the dark energy component, which will characterize the scaling of 
the expansion factor. We consider models where the present equation of state 
of the dark energy is $w_0 = -0.9$, consistently with the present constraints 
\cite{seljak_etal_2006}. The redshift evolution of $w(z)$ is shown in Fig. 
\ref{wred}, showing that while in the IPL model it is mildly departing from 
its present value at high redshifts, in the SUGRA one it rapidly gets to higher 
values. The different behavior of the dark energy density affects of course directly 
$H^{2}$, which is also plotted relatively to the two models in the figure; the difference 
peaks between redshift $1$ and $2$, and then quickly decreases, 
due to the increasing matter dominance. 
The remaining cosmological parameters are chosen accordingly to the concordance 
$\Lambda$CDM model, listed in the first column of table \ref{tablecosmpar}. \\
Since in the SUGRA model the dark energy keeps being relevant at higher 
redshifts with respect to the IPL case, the inhibition 
of structure formation starts earlier (see \cite{bartelmann_etal_2002} and 
references therein); thus, for a fixed primordial normalization, we expect two 
effects. The first is a smaller lensing signal in the SUGRA case, where clustering 
suppression starts earlier. The second is that the redshift interval where the lensing 
signal picks up its power, formally defined here below, is shifted towards 
earlier epochs, as a consequence of the structure formation process occurring 
at higher redshifts following the earlier dark energy dominance. \\
These features can be verified analyzing the redshift behavior of the function 
which appears in the integral defining $\sigma_{0}(\theta)$; 
we choose a reference value of the angle, say $\theta_0 = 4 \cdot 10^{-3}$ radians, 
corresponding roughly to the middle of the range suitable for CMB computations. 
We consider the function ${\it k}(\theta_0,z)$, which gives $\sigma_{0}^{2}$ in 
(\ref{sigma0}) when integrated over $z$, assuming an unitary power spectrum for the 
gravitational potential fluctuations. We call the resulting quantity the 
{\it lensing kernel}; its dimensions are the inverse of a volume, 
and it gives a measure of the redshift distribution of the lensing effect coming 
from the background cosmological expansion. The result is shown in Fig.~\ref{fig:geomSI}. 
Both our expectations are verified; note how the two cosmological models, 
although having the same values of all cosmological parameters today, 
differ substantially ($30\%$) at the peak entirely because of the behavior 
of the dark energy equation of state. The corresponding behavior for the 
function giving $\sigma_2^{2}$ when integrated over $z$, evaluated 
on the same scale and using an unitary power spectrum, is qualitatively 
similar; indeed, the relevant quantity is $W^2 = (1 - \chi/\chi_{\rm rec})^2$, 
appearing multiplied by functions vanishing at present in the integral of both 
(\ref{sigma0}) and (\ref{sigma2}). \\
Let us now turn to analyze the impact of the different 
perturbation growth rate, influencing $\sigma_{0}(\theta)$ and 
$\sigma_{2}(\theta)$ through the power spectrum of the gravitational 
potential. It is convenient to plot the linear growth factor, 
$g(\tau)$, for the two models at a fixed wavenumber; the 
behavior is qualitatively the same for any $\kappa$. The 
result is shown in Fig.~\ref{fig:tfdeltaSI}.
The phenomenology is the following. In the matter dominated era, say at 
redshifts between 1000 and a few in the figure, $g$ has the well known 
scaling as $a=1/(1+z)$. At the onset of acceleration, its growth is 
inhibited and eventually it starts to decrease. As expected, this 
effect is stronger in the SUGRA case, as dark energy dominance takes 
place earlier. We also notice that the relation between the curves in 
Fig. \ref{fig:tfdeltaSI} is similar to the ones in \ref{fig:geomSI}: 
the case of SUGRA has less power with respect to the IPL. This brings 
us to fix the quantity which matters here, i.e. causing the differences 
we just discussed. That is simply the dark energy abundance at the epoch 
in which lensing injects its power, corresponding to the interval outlined 
by the lensing kernel in Fig. \ref{fig:geomSI}. Indeed, the latter as well as 
the perturbation growth rate are determined by the Hubble expansion rate, 
which contains the dark energy through the Friedmann equation. The higher 
is the dark energy, the higher $H$, causing an higher suppression of 
perturbations, Fig.\ref{fig:tfdeltaSI}, as well as modified geometry, 
Fig.\ref{fig:geomSI}. Unfortunately, there is no unique parameter fixing 
the dark energy abundance at a given redshift, which is determined 
in general by the present abundance and $w(z)$; in our parametrization, 
usually the most relevant parameter is $w_{\infty}$, specifying the 
dark energy density redshift behavior at high redshifts, but only 
approximately in models where $w_{\infty}$ is sensibly larger than 
$w_{0}$. On the other hand, the improvement from having the lensing 
effect in the CMB analysis is evident keeping the parametrization we 
have exploited so far, as we show in the following. As a final remark of 
this part of the discussion, we notice that the the combination of the two 
effects of background evolution and perturbations growth contributing to the 
expressions of $\sigma_0$ and $\sigma_2$ is indeed large. We thus expect a 
significant dependence of the amplitude of CMB lensing power upon the dark energy 
equation of state value in the redshift interval which is relevant for lensing, 
i.e. the one outlined by the lensing kernel in Fig. \ref{fig:geomSI}. 

\begin{figure}
\includegraphics[width=\linewidth]{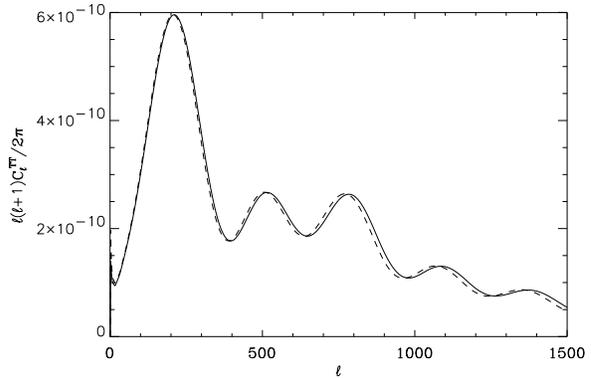} 
\caption{TT lensed power spectra for the SUGRA (dashed line) 
and IPL (solid line) models.}
\label{fig:clt}
\end{figure}

\begin{figure}
\includegraphics[width=\linewidth]{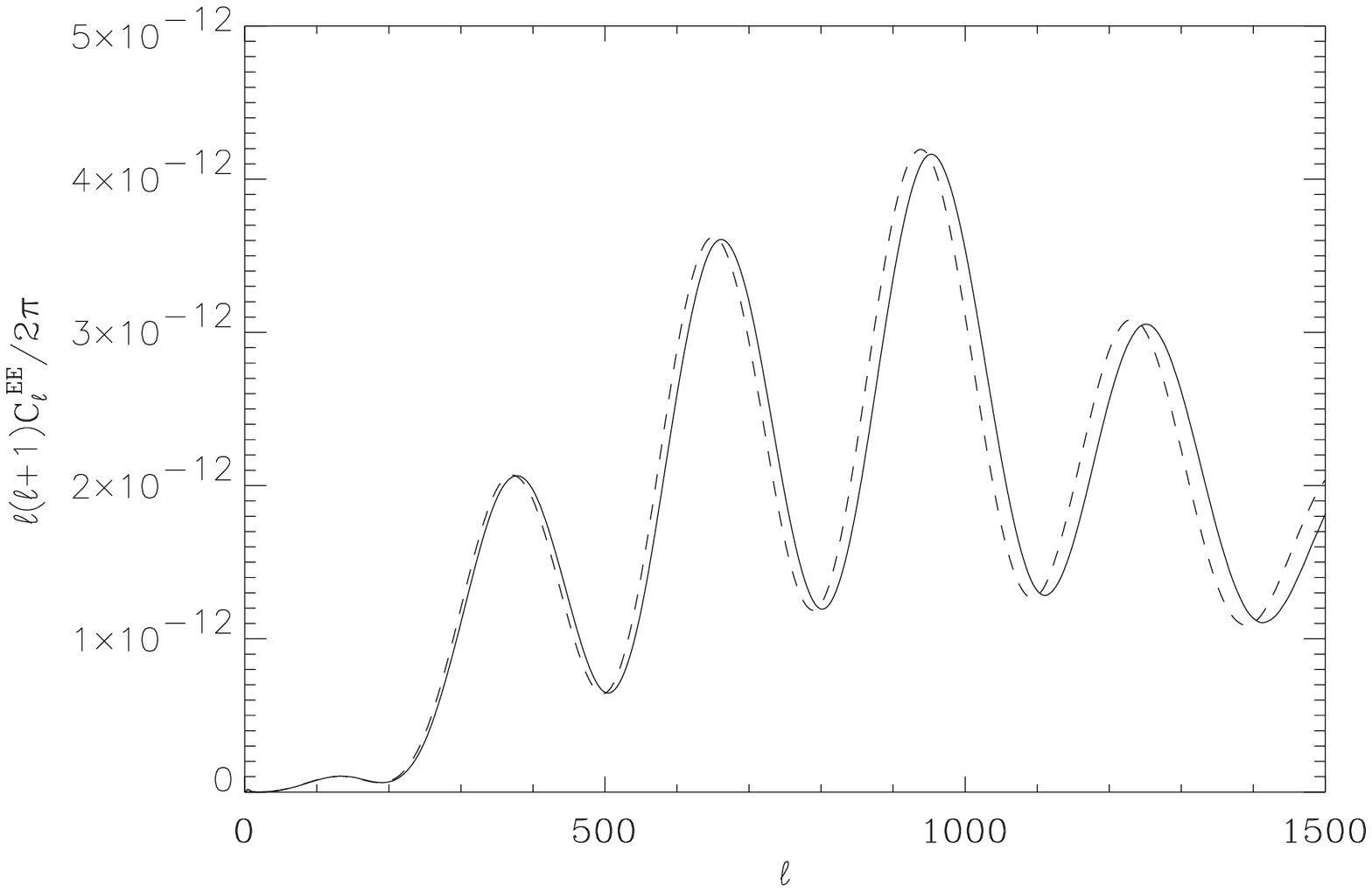} 
\caption{EE lensed power spectra for the SUGRA (dashed line) 
and IPL (solid line) models.}
\label{fig:cle}
\end{figure}

\begin{figure}
\includegraphics[width=\linewidth]{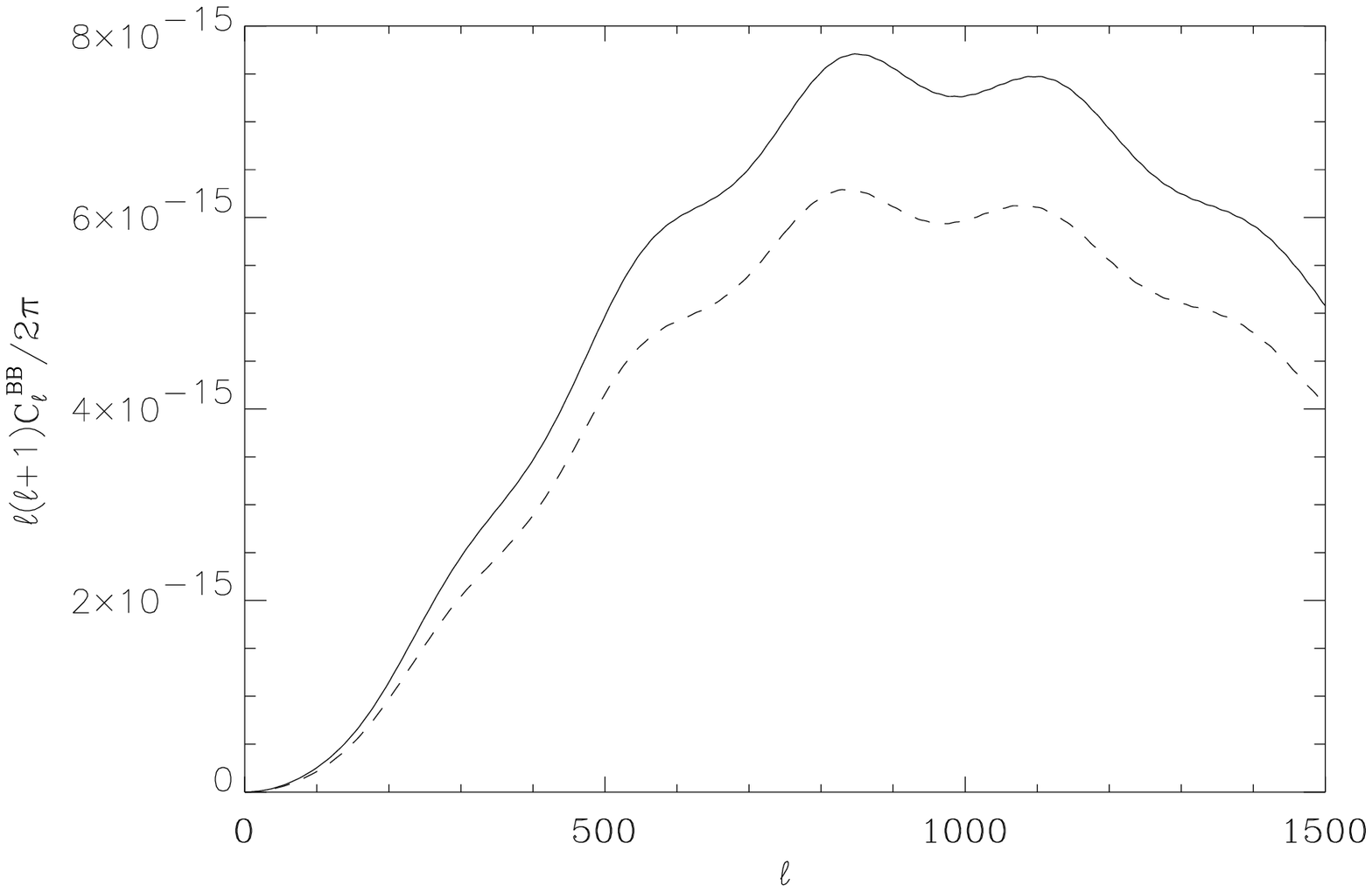} 
\caption{BB lensed power spectra for the SUGRA (dashed line) 
and IPL (solid line) models.}
\label{fig:clb}
\end{figure}

\begin{figure}
\includegraphics[width=\linewidth]{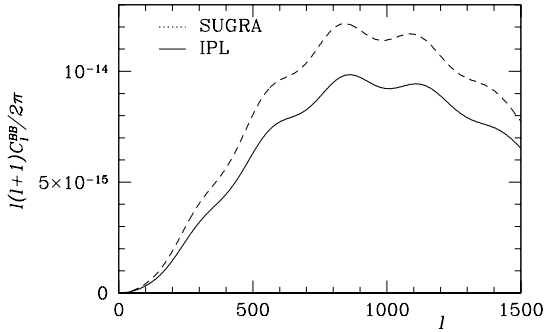} 
\caption{Lensed BB power spectra for the IPL and SUGRA models, 
normalized to have the same $\sigma_{8}$ at present.}
\label{fig:clbsigma8}
\end{figure}

On the basis of the issues outlined above, it is crucial to fix CMB 
observables purely sourced by gravitational lensing. 
The BB modes in the CMB represent an almost ideal candidate for this, 
since the lensing dominates their power on sub-degree angular scales. 
In the following we give a qualitative illustration of their relevance, leaving 
a more quantitative discussion for the next Section. \\ 
The TT, TE and EE CMB spectra are dominated by primary anisotropies, 
imprinted at last scattering, where in most models the dark energy 
is not yet effective. The location of the acoustic peaks depends on the different cosmological 
expansion histories, as a result of the modification in the comoving 
distance to last scattering $d_{LS}$, which is written as 
\bea
\label{dls}
&&d_{LS}= H^{-1}_0 \int^{z_{LS}}_0 dz[\Omega_m (1+z)^3+ 
\nonumber\\
&&+(1-\Omega_{m})e^{3\int^z_0 dz^{\prime}
\frac{1 + w(z^{\prime})}{1 + z^{\prime}}}]^{-1/2}\,. 
\eea
where $H_{0}$ is the Hubble parameter, $\Omega_{m}$ is the matter 
abundance today relative to the critical density and the contributions 
from radiation and curvature are neglected. 
It does not come as a surprise that this quantity depends very weakly on 
different forms of $w(z)$, since those are washed out by two integrals in 
redshift; the latter occurrence gives rise to the so-called 
projection degeneracy of CMB anisotropies. This simply comes from the 
fact that $d_{LS}$ is the same for different combinations of parameters in 
(\ref{dls}); in particular, for a given set of those except the dark energy 
equation of state, there is an entire set of curves $w(z)$ giving rise to 
the same $d_{LS}$; in general, in all those models the dark energy is negligible 
at decoupling, so that the shape of acoustic peaks is also unchanged, making the 
spectra in those models nearly identical, and therefore degenerate. 
The Integrated Sachs-Wolfe effect (ISW) acts on large scales only, responding 
to the change in the cosmic equation of state; although promising results 
may be obtained correlating the ISW with the large scale structure 
data \cite{gold_2005}, from a pure CMB point of view the cosmic 
variance represents a substantial limiting factor. \\
The BB phenomenology is utterly different. Here the lensing is the 
only source of power on sub-degree angular scales, and the 
lensing cross section is largest at intermediate redshifts, say around 
$z \simeq 1$ as Fig. \ref{fig:geomSI} shows, where 
the dark energy might differ significantly from a Cosmological 
Constant, even for the same expansion rate today. The lensed CMB power 
spectra are shown in Figs.~\ref{fig:clt},~\ref{fig:cle},~\ref{fig:clb}. 
The TT, TE and EE spectra undergo a projection effect due to the modified 
distance to last scattering, plus a lensing distortion which is barely visible, 
as the curves are still dominated by the primordial 
component of anisotropies. On the other hand, the BB spectrum
amplitude is markedly affected by the different lensing process in the two models. 
Indeed, as we discussed above, the BB peak directly traces the perturbation 
growth rate and the background expansion in the redshift interval visible in 
Fig. \ref{fig:geomSI}. For the cosmological parameters at hand, the 
effect is of 20-30 percent, consistently with 
the results in Figs.~\ref{fig:geomSI},~\ref{fig:tfdeltaSI}. We remark 
that the $C_{l}$s have been obtained with the same value of all 
cosmological parameters, including the primordial normalization, 
and differ only in the value of the dark energy equation of state 
at redshifts relevant for lensing. As we explained above, that is 
equivalent to a different dark energy abundance between the models 
considered at that epoch, and therefore a different expansion rate, 
which determines the strength of the lensing process. In order to 
check that the magnitude of the effect we point out does not depend 
on the normalization procedure adopted, in Fig. \ref{fig:clbsigma8} we 
plot the BB lensing peak for the SUGRA and IPL models, normalized 
to have the same $\sigma_{8}$ at present, which is chosen to be 
$0.844$. Correspondingly, the models need now to start from a different 
primordial normalization, in order to get the same present power, 
given their different perturbation growth histories represented 
in Fig. \ref{fig:tfdeltaSI}. As it is evident, the magnitude of the 
difference is comparable to the one in Fig. \ref{fig:clb}, but the 
order of the curve is reversed. This may be understood by looking again 
at Fig. \ref{fig:tfdeltaSI}; starting from the same power at present, 
and going towards higher redshifts, in the SUGRA model perturbations 
are higher at any epoch, determining the opposite behavior with respect 
to Fig. \ref{fig:clb}. Thus, the fact that the lensing difference is 
comparable in both cases makes us confident that the effect we point 
out is not an artefact of the normalization procedure. 

\begin{figure}
\includegraphics[width=\linewidth]{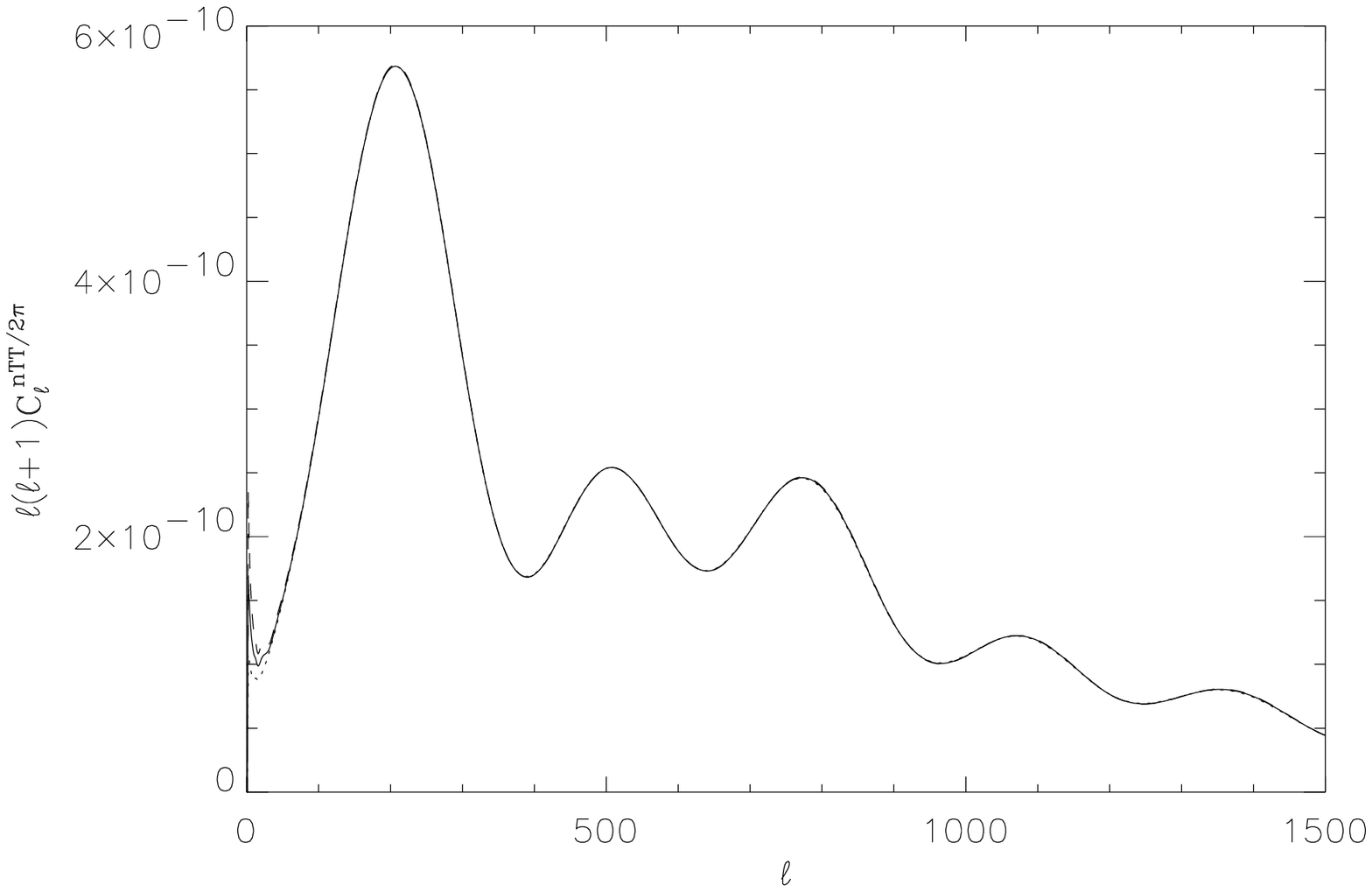} 
\caption{Lensed TT power spectra for dark energy models 
with $w_{0}=-0.9$, $w_{\infty}=-0.4$ (solid line), 
$w_{0}=-0.965$, $w_{\infty}=-0.3$ (dashed line), 
$w_{0}=-0.8$, $w_{\infty}=-0.56$ (dotted line).}
\label{fig:cltdls}
\end{figure}

\begin{figure}
\includegraphics[width=\linewidth]{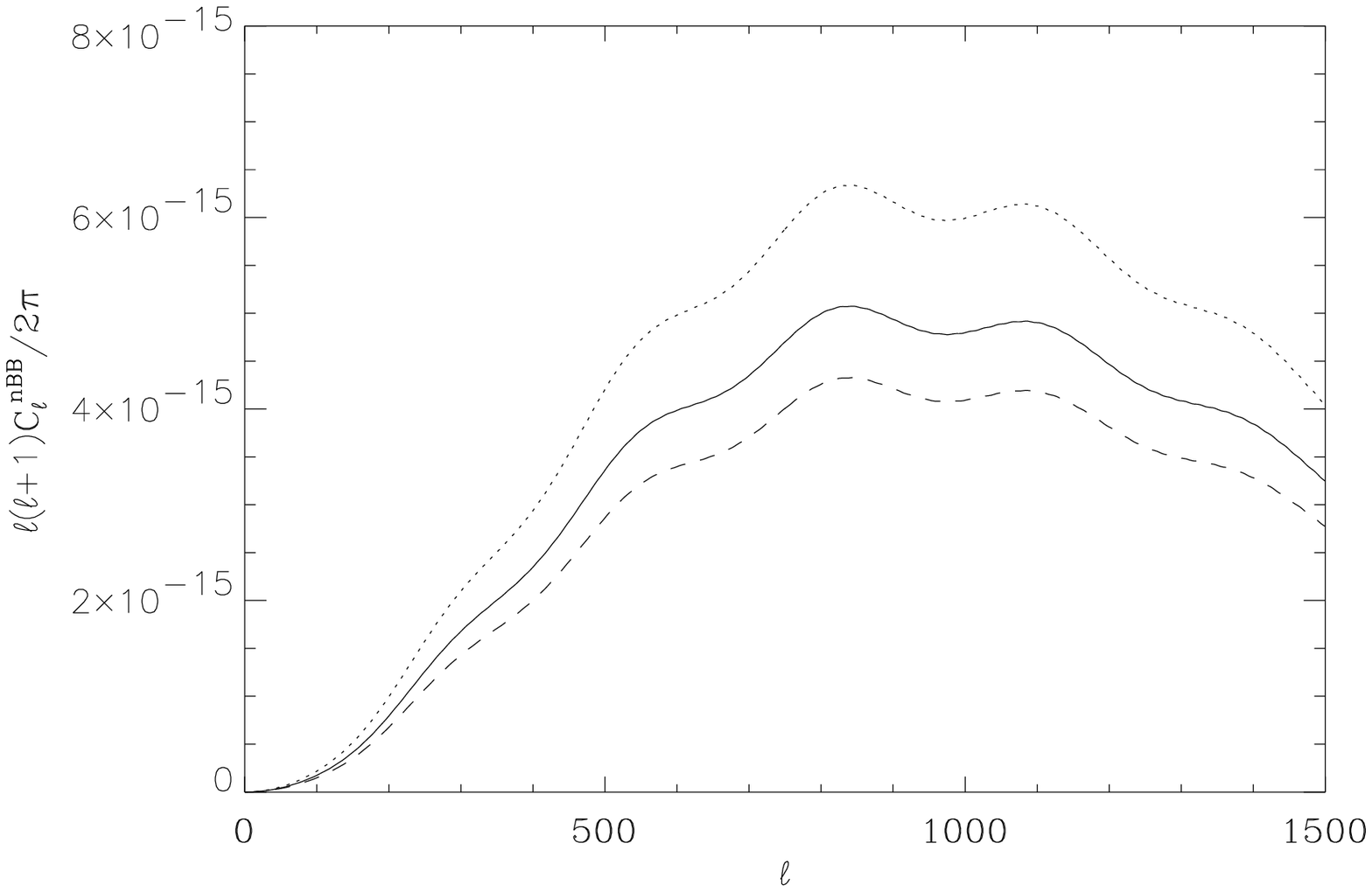} 
\caption{Lensed BB power spectra for dark energy models 
with $w_{0}=-0.9$, $w_{\infty}=-0.4$ (solid line), 
$w_{0}=-0.965$, $w_{\infty}=-0.3$ (dashed line), 
$w_{0}=-0.8$, $w_{\infty}=-0.56$ (dotted line).}
\label{fig:clbdls}
\end{figure}

We now demonstrate how the lensing breaks the projection 
degeneracy mentioned above. For simplicity, here and in the rest 
of this Section we adopt the simple parametrization in terms of 
$w_{0}$ and $w_{\infty}$. 
Let's consider dark energy models featuring the same value of $d_{ls}$ 
in (\ref{dls}), with different values of $w_{0}$ and $w_{\infty}$. 
The TT and BB spectra are shown in Figs. ~\ref{fig:cltdls} ,~\ref{fig:clbdls}, 
showing clearly the same pattern in the TT acoustic peaks but markedly 
different BB amplitude, reflecting the enhanced dependence of the 
latter on $w_{\infty}$. It is appropriate here to make a connection 
with the issue outlined before of the relevance of the dark energy 
abundance through $H$. 
The dark energy density at the epoch which is relevant for lensing, 
see Fig.\ref{fig:geomSI} again, follows an opposite behavior with 
respect to the curves represented in the figure: the lower the curve, 
the higher the value of the expansion rate at the lensing relevant epoch 
leading to an increasing suppression of the power, the higher the dark 
energy density at the corresponding redshifts, which is mainly influenced 
by $w_{\infty}$ here, as the present dark energy abundance is the same. 

\begin{figure}
\includegraphics[width=\linewidth]{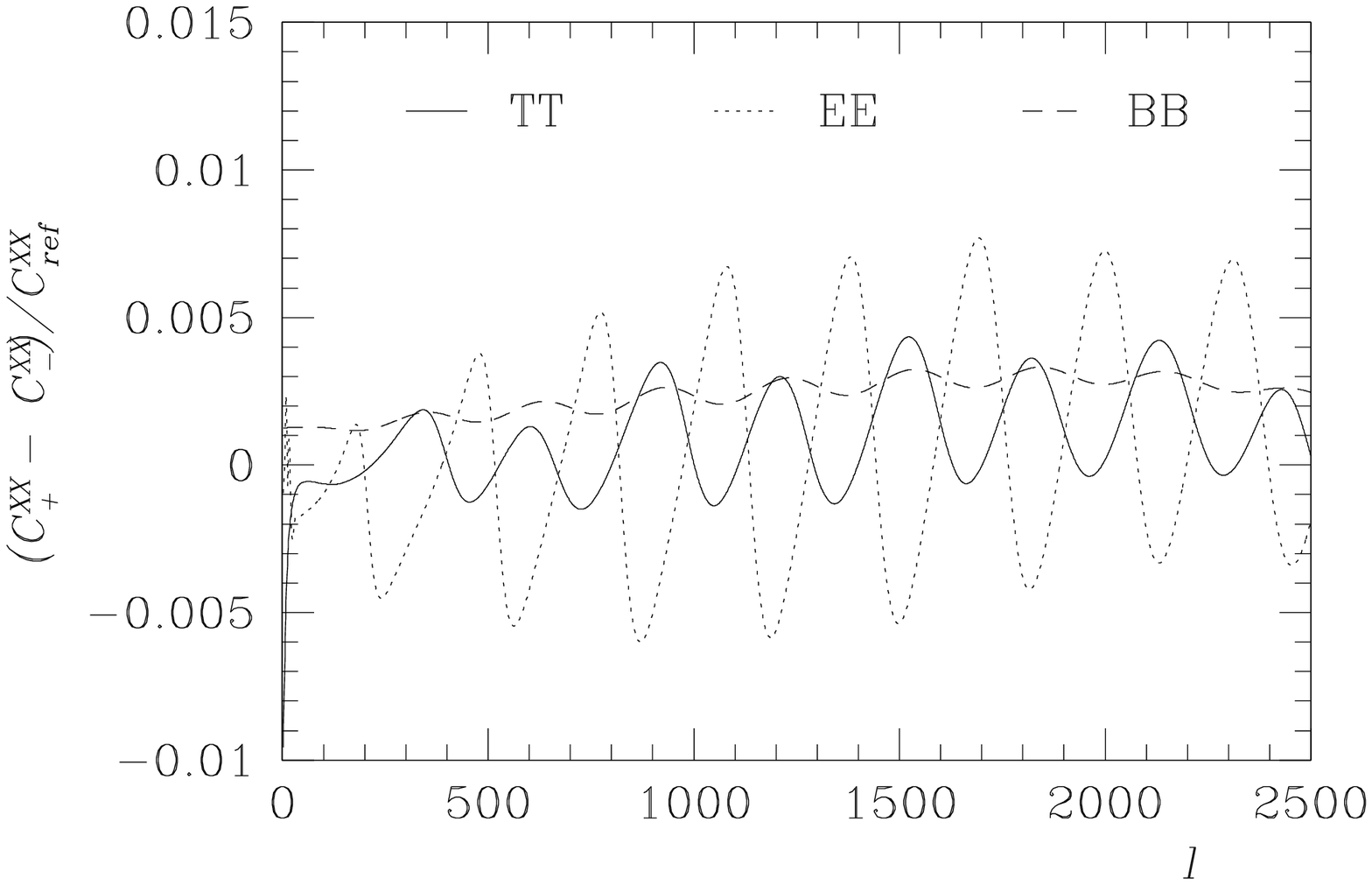} 
\caption{Relative lensing changes for the TT, EE and BB 
CMB spectra, varying around a $\Lambda$CDM cosmology.}
\label{fig:derivative_LAMBDA}
\end{figure}

Finally, we wish to address an important point which will be relevant in the 
next Section, i.e. the lensing distortion on the TT, TE and EE modes. Indeed, 
most of the reasoning exposed in this Section, represented by the phenomenology 
in Figs.  \ref{fig:geomSI} and \ref{fig:tfdeltaSI}, does not apply to the 
lensing BB modes only, but to every effect coming from lensing. It is therefore 
relevant to compare the lensing effect on non-BB modes, too. This is done 
in Fig. \ref{fig:derivative_LAMBDA}, where we plot the quantity 
$(C^{XX}(w_{\infty} +dw_{\infty}) - C^{XX}(w_{\infty}))/C^{XX}(w_{\infty})$, 
where XX stays for TT, EE and BB. Such quantity represents the fractional change to the 
spectra induced by a different dark energy abundance at the epoch in which the lensing 
is active, see Fig. \ref{fig:geomSI} again; the latter is determined entirely by 
$w_{\infty}$, as $w_{0}$ is fixed to $-1$, and the present dark energy abundance is 
the same. We perform a double-sided variation around the $\Lambda$CDM cosmology, so that 
$w_{\infty}=-1$, and $dw_{\infty=0.05}$. As it is evident, the changes have a comparable 
order of magnitude for all the spectra, including the BB one; remarkably, the latter is not oscillating 
around zero, but possesses a definite sign. This is due to the absence of sharp 
peaks and valleys in the spectrum, which in turn comes from the lensing capability 
of correlating different scales, smearing out the EE peaks which represent the 
source for BB lensing modes. That may be a relevant aspect for 
experiments looking at limited sky patches, for which a binning procedure is 
required. The binning would of course reduce the relevance of the lensing distortion 
on the TT and TE, while the BB would remain substantially unaffected. 

In the next Section, we will evaluate the relevance of considering the 
lensing effect for cosmological parameter estimation. 

\section{Fisher matrix analysis}
\label{fma}

Here we give a first quantitative evaluation of the benefit 
that the knowledge of the lensing and in particular the BB 
spectra has on the CMB capability of constraining the dark 
energy dynamics. Our approach is based on a Fisher matrix 
analysis, reviewed in Section \ref{met}; in Section \ref{meocp} 
we show the results. 

\subsection{Method}
\label{met}

In a CMB analysis involving the polarization power spectra 
\cite{zaldarriaga_seljak_1997}, the Fisher matrix takes the form 
\begin{equation}
\label{eq:fishc} {\rm F}_{ij}=\sum\limits_{\ell}\sum\limits_{XY}  
\frac{\partial C_{\ell}^{X}}{\partial \alpha_i} \left[\,
\Xi_{\ell}\right]_{XY}^{-1} \frac{\partial C_{\ell}^{Y}}{\partial 
\alpha_j},
\end{equation}  
where $X$ and $Y$ are either TT, EE, TE or BB and $  \Xi_{XY}
\equiv {\rm Cov}(C^X_\ell C^Y_\ell) $ is the power spectra
covariance matrix: 
\begin{equation}
\label{eq:PScov} {\Xi}_{\ell}= \left(\begin{array}{cccc}
{\Xi}_{\ell}^{TT,TT} & {\Xi}_{\ell}^{TT,EE} & {\Xi}_{\ell}^{TT,TE} & 0 \\
{\Xi}_{\ell}^{TT,EE} & {\Xi}_{\ell}^{EE,EE} & {\Xi}_{\ell}^{EE,TE} & 0 \\
{\Xi}_{\ell}^{TT,TE} & {\Xi}_{\ell}^{EE,TE} & {\Xi}_{\ell}^{TE,TE} & 0 \\
0 & 0 & 0 & {\Xi}_{\ell}^{BB,BB}
\end{array}\right)\ .
\end{equation}
The terms in the power spectra covariance matrix are given by 
\begin{eqnarray}
\label{eq:XYXY}
\Xi_{\ell}^{xy,x'y'}&=&\frac{1}{(2\ell+1)f_{sky}\Delta \ell} \nonumber \\
& \times& \big[(C_{\ell}^{xy'}+N_{\ell}^{xy'})
(C_{\ell}^{yx'}+N_{\ell}^{yx'})\nonumber \\
& & + (C_{\ell}^{xx'}+N_{\ell}^{xx'})   
(C_{\ell}^{yy'}+N_{\ell}^{yy'})\big],
\end{eqnarray}
where $(x,y)=(T,E,B)$. 
The noise covariance is given by $N_{\ell}^{xy}$, which also 
contains the effect of the instrumental beam, assumed Gaussian 
and circular. The inverse of the Fisher matrix gives the uncertainty 
on the theoretical parameters: 
\begin{equation}
\label{covariancematrix}
{\cal C}_{ij} \equiv \left< \Delta \alpha_i \Delta \alpha_j \right>
={\rm F}_{ij}^{-1}\ . 
\end{equation}
$\Delta \alpha_i$ is the marginalized 1-$\sigma$ error on the 
$i^{\rm th}$ parameter, and is given by the square root of 
the diagonal elements of the inverse of the Fisher matrix. \\
As a representative of the forthcoming CMB polarization probes 
capable to detect the BB spectrum we consider a post-Planck all-sky experiment. 
We conservatively consider a Gaussian beam with 
7 arcminutes full width half maximum, considering multipoles up to 
$l=1800$. We assume an instrumental error of 1 $\mu$ K on the beam scale, 
and cut the galactic plane assuming a sky fraction of 0.66. \\
A delicate issue in applying a Fisher matrix analysis to the 
CMB lensing is represented by the non-Gaussianity of the lensing 
effect, due to the correlation of cosmological perturbations on 
different angular scales; the lensing statistics is being investigated, 
receiving increasing attention in view of the incoming precision polarization 
experiments \cite{smith_etal_2006,lewis_2005,smith_etal_2004,kaplinghat_etal_2003,hirata_seljak_2003}.\\
In particular, Smith et al. \cite{smith_etal_2004} achieved a first 
quantification of the increase in the covariance matrix due to the 
non-Gaussian nature of the lensing signal in the BB modes, giving a pipeline
to estimate the resulting achievable accuracy. \\
For our study their most relevant result is the behavior of the
so called {\it degradation factor}, the ratio between the squared sample
covariance in the case of this non Gaussian signal and the corresponding
Gaussian case. This is shown to depend both on the instrumental error (the degradation
increases with the signal-to-noise ratio of the experiment, as expected because the
instrumental error is close to Gaussian) and on the maximum available multipole (again increasing 
with $l_{\text max}$, because of the stronger effect of the correlation between neighboring band 
powers). \\
According to their worst case scenario, we make a conservative choice 
enlarging by a factor $10$ the covariance contribution to the 
BB spectrum in the covariance matrix (\ref{eq:XYXY}); this would correspond to
an experiment with $l_{\text max} = 2000$ and $\theta_{\text FWHM} \simeq 1'$. \\
Steps further in the issue of taking into account has been made in \cite{smith_et_al}, 
who suggested a way of taking into account the non-Gaussian correlations of the lensed BB 
spectra,with special regard on the issue of degeneracies between the dark energy parameters 
and the neutrino mass, and most recently in \cite{smith_etal_2006}.  These approaches go however beyond the scope of the present paper, 
and may be considered in further work. 
\begin{table}[htb]
\begin{center}
\vspace*{0.3cm}
\caption{Results from the Fisher matrix analysis for the $\Lambda$CDM and the IPL models.}\label{tablecosmpar}
\vspace*{0.3cm}
\begin{tabular}{lcc|cc}
\hline \hline & \multicolumn{2}{c|}{\bf $\Lambda$CDM} &
\multicolumn{2}{c}{\bf IPL}\\
& value & $\sigma_{Fisher}$& value & $\sigma_{Fisher}$\\
\hline
${ w_{0} }$ & $- 1.$ & $0.12$
& $- 0.9$ & $9.7 \times 10^{-2} $ \\
${ w_{\infty} }$ & $ - 1.$ & $0.27$
&$- 0.8 $ & $ 0.19$ \\
 ${ \Omega_{b} h^2}$ &  $0.022$ & $5.7 \times 10^{-5}$
&  $0.022$ & $ 6.0 \times 10^{-5}$ \\
${ \Omega_{C} h^2}$ & $0.12$ & $7.0 \times 10^{-4}$
 & $0.12$ & $7.3 \times 10^{-4} $\\
${ h }$ & $0.72$ & $5.0 \times 10^{-2}$
& $0.72$ & $4.5 \times 10^{-2} $ \\
${ n_S }$ & $0.96$ & $2.1 \times 10^{-3}$
& $0.96$ & $2.2 \times 10^{-3}$ \\
${ \tau}$ & $0.11$ & $3.1 \times 10^{-3}$
& $0.11$ & $3.0 \times 10^{-3}$\\
${ A_S }$ & $1.0$ & $5.6 \times 10^{-3}$ &
$1.0$ & $5.5 \times 10^{-3}$ \\
 \hline
\end{tabular}
\end{center}
\end{table}

\begin{table}[htb]
\begin{center}
\vspace*{0.3cm}
\caption{Results from the Fisher matrix analysis for two SUGRA models.}\label{tablecosmpar2}
\vspace*{0.3cm}
\begin{tabular}{lcc|cc}
\hline \hline & \multicolumn{2}{c|}{\bf SUGRA1} &
\multicolumn{2}{c}{\bf SUGRA2}\\
 & value & $\sigma_{Fisher}$& value & $\sigma_{Fisher}$\\
\hline
${ w_{0} }$ & $- 0.9$ & $6.1 \times 10^{-2}$
& $- 0.82$ & $3.5 \times 10^{-2}$ \\
${ w_{\infty} }$ & $ - 0.4$ & $6.9 \times 10^{-2}$
&$- 0.24 $ & $ 1.9 \times 10^{-2}$ \\
 ${ \Omega_{b} h^2}$ &  $0.022$ & $5.7 \times 10^{-5}$
&  $0.022$ & $ 5.9 \times 10^{-5}$ \\
${ \Omega_{C} h^2}$ & $0.12$ & $6.6 \times 10^{-4}$
 & $0.12$ & $5.0 \times 10^{-4} $\\
${ h }$ & $0.72$ & $2.9 \times 10^{-2}$
& $0.72$ & $1.5 \times 10^{-2} $ \\
${ n_S }$ & $0.96$ & $2.1 \times 10^{-3}$
& $0.96$ & $2.0 \times 10^{-3}$ \\
${ \tau}$ & $0.11$ & $3.1 \times 10^{-3}$
& $0.11$ & $3.2 \times 10^{-3}$\\
${ A_S }$ & $1.0$ & $5.5 \times 10^{-3}$ &
$1.0$ & $5.6 \times 10^{-3}$ \\
 \hline
\end{tabular}
\end{center}
\vspace*{2cm}
\end{table}

\subsection{Marginalized errors on cosmological parameters}
\label{meocp}

We analyze four cosmological models, corresponding to a pure $\Lambda$CDM, and inverse power law, 
and two SUGRA cases, specified by eight cosmological parameters, including the two 
specifying the dark energy equation of state: 
\begin{tabbing}
\hspace*{0.5cm} \= $w_0$~~~~~ \= present e.o.s. of dark energy\\
\> $w_\infty$ \> asymptotic past e.o.s. of dark energy \\
\> $h$ \> present value of Hubble parameter \\
\> $\Omega_{{\rm B}} h^2 $\> fractional baryon density $\times \; h^2 $\\
\> $\Omega_{{\rm C}} h^2 $ \> fractional CDM density $\times \; h^2$\\
\> $A_{{\rm S}}$ \> primordial normalization parameter\\
\> $n_{{\rm S}}$ \> perturbation spectral index\\
\> ${\tau}$ \> reionization optical depth \\
\end{tabbing}
All the cosmological parameters, are chosen consistently with the current 
observations of CMB and large scale structure \cite{seljak_etal_2006}; of course 
the dark energy equation of state is allowed to depart from a $\Lambda$CDM case. 
We assume the same values for the non-dark-energy parameters for all the models. 
Their values are listed in the first and third column of the Tabs. \ref{tablecosmpar}, 
\ref{tablecosmpar2}. As we specified in the previous Section, in our numerical 
machinery the normalization is performed through an input parameter specifying 
the primordial power. For all the four reference models we run DEfast with the in-built
large-scale normalization option (COBEnormalize), we correct in order to reproduce the
best fit of the combined datasets of WMAP 1st year, CBI and ACBAR (see \cite{spergel_etal_2003} and references therein),
 and we use these four numbers
as the reference primordial amplitudes for each model. The parameter $A_S$ is the ratio of the 
primordial amplitude with respect to the latter; this is done for notational convenience since
the units of DEfast, based on version 4.1 of CMBfast, are not easily interpreted in terms of physical quantities.
The value $A_S = 1$, reported for the four cases, does not therefore indicate that the models have the same
amount of primordial perturbations, but simply that for each case the adopted normalization
is the one obtained with the procedure described above, whose actual value is, of course, model-dependent. Indeed
all the models give rise to a similar value for $\sigma_8$, which is $\simeq$ 0.86 for the $\Lambda$CDM case
and $\simeq$ 0.81 for the highly dynamical SUGRA2.\\
The results of the analysis are shown in the second and fourth column of 
tables \ref{tablecosmpar}, \ref{tablecosmpar2}, reporting the $1-\sigma$ marginalized 
errors for each parameter, according to the present Fisher matrix approach. 
For the $\Lambda$CDM model there is an important indication of an achievable precision smaller 
than $20\%$ on the $w_0$ parameter, while the limit on $w_\infty$ is considerably weaker. 
The accuracy on the others is in agreement with previous similar analysis \cite{balbi_etal_2003}, which 
was indeed expected because the BB modes statistics have large error bars and trace the physics at late 
redshifts so that their influence on other parameters is smaller. Results are increasingly better for 
the IPL and the two SUGRA cases; this can be attributed to the more and more violent redshift behavior 
of the equation of state of these models, making them increasingly sensitive to the redshift region probed 
by lensing, outlined in Fig. \ref{fig:geomSI}. In particular, the achievable precision on the $w_\infty$ 
parameter appears to be growing faster, so that for the SUGRA cases the results for the two dark energy 
parameters are comparable. \\
As we did in the discussion in the last part of the previous Section, it is relevant to evaluate 
the role of the BB modes compared to the lensing distortion on the TT, TE and EE spectra. 
When the lensing effect is not considered, i.e. performing an analysis only 
on the unlensed ${\cal C}^{TT}$,  ${\cal C}^{TE}$ and ${\cal C}^{EE}$ spectra, 
the projection degeneracy related to the last scattering surface distance 
(\ref{dls}) is almost exact making in particular the Fisher matrix singular. When 
the lensing is taken into account, even on the spectra which are actually 
dominated by the primordial power, namely the TT, TE and EE ones, such 
degeneracy is broken, as Fig. \ref{fig:derivative_LAMBDA} proves. It is 
interesting to compare the forecast precision on the cosmological 
parameters we consider, in particular $w_{0}$ and $w_{\infty}$, 
in presence of absence of the lensing BB modes. Looking at Fig. 
\ref{fig:derivative_LAMBDA}, we roughly expect a precision increase of 
order $20\%$ in the $w_{\infty}$ parameter, as the order of magnitude of the 
relative variation in the BB power is of the order of the other ones. The results 
are shown in tables \ref{tab:withorwithoutB1} and \ref{tab:withorwithoutB2} and 
confirm our expectation, in some cases being even larger than the naive expectation due
to the peculiar sensitivity of the BB modes spectrum to the dark energy equation of state
derivative.
 It is also interesting to note that there are significant improvements 
in the precision on some of the remaining parameters as well, as a result of the addition 
of a new independent observable. \\
In the next Section we further comment these results, 
and draw our conclusions. 
\begin{table}[htb]
\begin{center}
\vspace*{0.3cm}
\caption{Results from the Fisher matrix analysis in absence or presence of the BB modes.}
\label{tab:withorwithoutB1}
\vspace*{0.3cm}
\begin{tabular}{lcc|cc}
\hline \hline &  \multicolumn{2}{c|} {\bf $\Lambda$CDM} &
\multicolumn{2}{c} {\bf IPL} \\
 &  $\sigma$ (no BB)&  $\sigma$ (with BB) &  $\sigma$ (no BB) &  $\sigma$ (with BB) \\
\hline
 $ { w_{0} }$ &  $0.13$ &  $0.12$
&  $0.11$ &  $9.7 \times 10^{-2} $ \\
 ${ w_{\infty} }$ &  $ 0.31 $ &  $0.27$
& $0.24 $ & $ 0.19$ \\
 ${ \omega_B}$ & $6.4 \times 10^{-5}$ & $5.7 \times 10^{-5}$
&  $6.5 \times 10^{-5}$ & $ 6.0 \times 10^{-5}$  \\
 ${ \omega_C}$ & $7.9 \times 10^{-4}$ & $7.0 \times 10^{-4}$
 & $7.8 \times 10^{-4}$ & $7.3 \times 10^{-4} $ \\
${ h }$ & $5.6 \times 10^{-2}$ & $5.0 \times 10^{-2}$
& $5.4 \times 10^{-2}$ & $4.5 \times 10^{-2} $ \\
 ${ n_S }$ & $2.3 \times 10^{-3}$ & $2.1 \times 10^{-3}$
& $2.3 \times 10^{-3}$ & $2.2 \times 10^{-3}$ \\
${ \tau}$ & $3.2 \times10^{-3}$ & $3.1 \times 10^{-3}$
& $3.0 \times 10^{-3}$ & $3.0 \times 10^{-3}$ \\
${A_S}$ & $5.7 \times 10^{-3}$ & $5.6 \times 10^{-3}$ &
$5.6 \times 10^{-3}$ & $5.5 \times 10^{-3}$ \\
 \hline
\end{tabular}
\end{center}
\end{table}

\begin{table}[htb]
\begin{center}
\vspace*{0.3cm}
\caption{Results from the Fisher matrix analysis in absence or presence of the BB modes.}
\label{tab:withorwithoutB2}
\vspace*{0.3cm}
\begin{tabular}{lcc|cc}
\hline \hline &  \multicolumn{2}{c|} {\bf SUGRA1} &
\multicolumn{2}{c} {\bf SUGRA2} \\
 &  $\sigma$ (no BB)&  $\sigma$ (with BB) &  $\sigma$ (no BB) &  $\sigma$ (with BB) \\
\hline
 $ { w_{0} }$ &  $0.6.6 \times 10^{-1}$ &  $6.1 \times 10^{-2}$
&  $3.7 \times 10^{-2}$ &  $3.5 \times 10^{-2} $ \\
 ${ w_{\infty} }$ &  $ 7.9 \times 10^{-2} $ &  $6.9 \times 10^{-2}$
& $ 2.1 \times 10^{-2} $ & $ 1.8 \times 10^{-2}$ \\
 ${ \omega_B}$ & $6.4 \times 10^{-5}$ & $5.7 \times 10^{-5}$
&  $6.5 \times 10^{-5}$ & $ 5.9 \times 10^{-5}$  \\
 ${ \omega_C}$ & $7.5 \times 10^{-4}$ & $6.6 \times 10^{-4}$
 & $7.0 \times 10^{-4}$ & $5.0 \times 10^{-4} $ \\
${ h }$ & $3.3 \times 10^{-2}$ & $2.9 \times 10^{-2}$
& $1.6 \times 10^{-2}$ & $1.5 \times 10^{-2} $ \\
 ${ n_S }$ & $2.2 \times 10^{-3}$ & $2.1 \times 10^{-3}$
& $2.3 \times 10^{-3}$ & $2.0 \times 10^{-3}$ \\
${ \tau}$ & $3.2 \times10^{-3}$ & $3.1 \times 10^{-3}$
& $3.2 \times 10^{-3}$ & $3.2 \times 10^{-3}$ \\
${ A_S }$ & $5.8 \times 10^{-3}$ & $5.5 \times 10^{-3}$ &
 $6.0 \times 10^{-3}$ & $5.6 \times 10^{-3}$ \\
 \hline
\end{tabular}
\end{center}
\end{table}

\section{Conclusions}
\label{c}
Our aim in this paper is to study the potentiality of the CMB physics, with
particular regard to the BB modes of the polarization, which are sourced 
by gravitational lensing of cosmic structures, in order to constrain the 
dark energy dynamics at the epoch of equivalence with the non-relativistic 
matter component. We focus on the lensing effect and in particular on the 
amplitude of the BB angular power spectrum; BB mapping techniques isolating the 
lensing power \cite{hirata_seljak_2003} might also be considered for extracting 
information about the cosmic expansion rate redshift behavior. We have shown how 
the CMB lensing, being directly linked to the cosmic dynamics and linear perturbation 
growth rate when the darkenergy enters the cosmic picture, presents an enhanced 
sensitivity to the value of the dark energy equation of state at the corresponding epoch. 
Such a feature breaks the so called projection degeneracy, affecting the unlensed spectra, 
preventing the possibility of constraining the redshift dependence of the dark energy 
equation of state from CMB. These features are particularly evident by looking at the 
response of the amplitude of the BB angular power spectrum induced by lensing; the 
latter is lowered by a factor as large as $30\%$ if the equation of state of 
the dark energy at high redshifts is raised to the value of typical Quintessence 
models, currently allowed by observations; correspondingly, the TT, TE, EE spectra 
undergo an angular shift in the acoustic peaks location, and a variation in the 
the smearing of acoustic peaks because of lensing. The reasons are that, on one hand, 
the lensing probes only intermediate redshifts between source and observers, and on 
the other, that the lensing dominates the BB power. The first aspect 
is clearly not specific to the CMB angular power spectrum, but may applied to any 
CMB lensing observable. Analogous studies have been focused on the non-Gaussian 
power injection into the anisotropy statistics of order larger than the second 
\cite{verde_spergel_2002,giovi_etal_2005}. Indeed, the outcome of these studies 
is consistent with the present one, i.e. the lensing power in the CMB bispectrum, 
the harmonic space analogue to the three point correlation function, presents 
a remarkable sensitivity to the dark energy equation of state at the onset 
of acceleration. Thus the present study is related to those, although on a 
completely different domain. Since we still don't know where the impact of 
instrumental systematics and foregrounds will be the strongest in a real 
experiment attempting to detect the CMB lensing signal, it is important to carry 
out the analysis on all CMB lensing observables, and in particular on the angular 
power spectrum. Our results show that the relative changes, $\delta C_{l}^{XY}/C_{l}^{XY}$ where 
$XY$ stays for TT, EE, TE or BB, induced by lensing are all of the same order of magnitude, 
but in addition there are at least two reasons why the BB signal should be taken 
into account. First, the BB modes in CMB polarization at the arcminute scale are the 
explicit target of forthcoming CMB probes (see ie \cite{oxley_etal_2005} and references therein). 
Second, their response to the variation of the dark energy abundance 
has a definitee sign, while the others oscillate around zero; this might cause 
a difference in favour of the lensing BB modes for experiments targeting a limited 
fraction of sky, due to the potential loss of information involved in the
binning procedure. \\
We have then quantified the scientific impact of our result in terms  
of the achievable precision on the cosmological parameters, evaluated through a 
Fisher matrix analysis, modeling our assumptions on the specifics of forthcoming 
probes of CMB polarization, for Cosmological Constant and three more dynamical dark energy 
scenarios. The results are strongly encouraging, predicting an accuracy better of order $10\%$ 
on the present value of the dark energy equation of state, and a somehow weaker limit on its 
first derivative with respect to the scale factor, but with an important indication of better 
results with increasing dark energy dynamics. The inclusion of the BB spectra is responsible 
at the 10 to 20 percent level for the quoted forecasts. This result is comparable with the one quoted in
\cite{Yeche_et_al}, where the authors take into account SNIa data and CMB physics 
but do not include BB modes into the analysis and with the forecasts in \cite{SNAP} for 
Quintessence models, where the authors consider SNIa data and weak lensing of background 
galaxies. In particular, the prediction of a smaller uncertainty for high dark energy 
dynamics is reproduced also for the observable considered here. The latter 
might be a complementary and independent dark energy probe with respect to the 
ones mentioned above. In \cite{hu_etal_2006}, a work which came out in the literature 
during the refereeing process of the present one, the lensing BB modes are included in the 
analysis, finding a benefit which is similar to the one found here. \\
In conclusion, the weak lensing of the CMB is confirmed by this work as a potential 
probe of the dark energy dynamics when acceleration starts, independently 
of the present expansion rate. This adds even more interest to the impact 
of high precision weak lensing measurements in cosmology 
\cite{refregier_2003}. Moreover, our results indicate that the measure of 
the dark energy dynamics suggested here could be achieved by the 
forthcoming CMB probes aiming at the detection of the polarization BB modes. 

\section{Acknowledgments}
\label{ac}
We thank Julien Lesgourgues, Samuel Leach, Matias Zaldarriaga and Matthias Bartelmann for useful 
conversations and suggestions. This research was in part supported by
the NASA LTSA grant NNG04GC90G.

\end{document}